# Visualizing nematic transition and dramatic suppression of superconductivity in Fe(Te,Se)

He Zhao[1], Hong Li[1], Lianyang Dong[2], Binjie Xu[3], John Schneeloch[4], Ruidan Zhong[4], Minghu Fang[3], Genda Gu[4], John Harter[2], Stephen D. Wilson[2], Ziqiang Wang[1] and Ilija Zeljkovic[1]

*[1] Department of Physics, Boston College, 140 Commonwealth Ave, Chestnut Hill, MA 02467*

*[2] Materials Department, University of California, Santa Barbara, CA 93106, USA.*

*[3] Department of Physics, Zhejiang University, Hangzhou 310027, China*

*[4] Brookhaven National Laboratory, Upton, New York 11973, USA*

**The interplay of different electronic phases underlies the physics of unconventional superconductors. One of the most intriguing examples is a high-$T_c$ superconductor FeTe$_{1-x}$Se$_x$ – it undergoes both a topological transition, linked to the electronic band inversion, and an electronic nematic phase transition, associated with rotation symmetry breaking, around the same critical composition $x_c$ where superconducting $T_c$ peaks. At this regime, nematic fluctuations and symmetry-breaking strain could have an enormous impact, but this is yet to be fully explored. Using spectroscopic-imaging scanning tunneling microscopy, we study the electronic nematic transition in FeTe$_{1-x}$Se$_x$ as a function of composition. Near $x_c$, we reveal the emergence of electronic nematicity in nanoscale regions. Interestingly, we discover that superconductivity is drastically suppressed in areas where static nematic order is the strongest. By analyzing atomic displacement in STM topographs, we find that small anisotropic strain can give rise to these strongly nematic localized regions. Our experiments reveal a tendency of FeTe$_{1-x}$Se$_x$ near $x$~0.45 to form puddles hosting static nematic order, suggestive of nematic fluctuations pinned by structural inhomogeneity, and demonstrate a pronounced effect of anisotropic strain on superconductivity in this regime.**

Fe-based high-$T_c$ superconductors have emerged as one of the leading platforms to study various correlated electron states and electronic phenomena associated with non-trivial topology. A prototypical example in this regard is superconductor FeTe$_{1-x}$Se$_x$ (Fe(Te,Se)) [1–11], which has been intensely investigated in recent years, predominantly motivated by the discovery of topological surface states [3,4] and Majorana zero modes [5,8,9,11] near critical composition $x_c$~0.45. A somewhat overlooked aspect of the intricate physics of Fe(Te,Se) near the same composition is that it is also expected to undergo an electronic nematic transition characterized by a rotation symmetry breaking, where the superconducting $T_c$ at ambient conditions reaches its peak [12,13]. At this regime, elastoresistance experiments uncovered signatures of nematic critical fluctuations and demonstrated an enormous impact of strain [14]. However, the implications of electronic nematicity and strain on the underlying physics of Fe(Te,Se) have not been fully explored to-date.

Experimental signatures of electronic nematicity in Fe-based superconductors include resistivity anisotropy typically accompanied by a small orthorhombic distortion below structural transition temperature $T_s$ [14–16], lifting of the band degeneracy [17] and a pronounced $C_2$-symmetric electron scattering [10,18–21]. A material can be tuned towards the nematic transition point by various parameters, such as chemical composition change [17,22–24], strain [14,16,25] and pressure [26,27]. Although the order parameter is typically considered only in the average crystal structure, it is conceivable that short range or local order may persist beyond this point. This behavior could emerge in a system close to the nematic phase transition, such as Fe(Te,Se) at $x_c$, which, although not in the orthorhombic phase at low temperature [12,13], shows signatures of nematic fluctuations [14] that can in principle get pinned to become static in localized regions.

We use low-temperature spectroscopic-imaging scanning tunneling microscopy (SI-STM) to study UHV-cleaved bulk single crystals of Fe(Te,Se) as a function of composition $x$ across the electronic nematic transition (Fig. 1(b)). Each unit cell of Fe(Te,Se) consists of an Fe layer sandwiched between two Se/Te chalcogen layers (Fig. 1(a)). STM topographs show a square atomic lattice with $a_0$~3.8 Å. The bright (dark) atoms in STM topographs represent Te (Se) atoms in the topmost layer (Fig. 1(c)), which could be counted to confirm the chemical composition (Supplementary Information 1). To gain a comprehensive insight into the evolution of electronic properties with $x$, we study Fe(Te,Se) samples with three different Se:Te bulk compositions: $x$~0.35, $x$~0.45 and $x$~0.5 (Fig. 1). Characteristic d$I$/d$V$ spectra for all three compositions show a superconducting gap of comparable size $\Delta_{SC}$ ~ 2 meV (insets in Fig. 1(d-f)), which likely represents a combination of the gaps of the hole pockets at Γ, all with similar magnitudes within ~0.5 meV [4,28].

In contrast to the relative insensitivity of the gap size on the composition $x$, normalized differential conductance maps L($\mathbf{r}$,$V$) (defined as d$I$/d$V$($\mathbf{r}$,$V$)/($I$($\mathbf{r}$,$V$)/$V$)), where $\mathbf{r}$ is the lateral tip position and $V$ is the bias applied to the sample) exhibit a remarkable evolution within the same composition range (Fig. 1(d-f)). Starting with the highest composition studied ($x$~0.5), we find that L($\mathbf{r}$,$V$) maps show nanometer-scale electronic modulations preferentially oriented along an Fe-Fe direction everywhere on the sample (Fig. 1(f), Fig. 2(j), Fig. S13). As we elaborate on in detail in subsequent paragraphs, this suggests that Fe(Te,Se) at $x$~0.5 is in an electronic nematic state. In contrast, L($\mathbf{r}$,$V$) maps of the lowest composition studied ($x$~0.35) show no sign of unidirectional features, indicative of a tetragonal state and the absence of electronic nematicity (Fig. 1(d), Fig. 2(i), Fig. S12). Interestingly, at the intermediate composition ($x$~0.45), we observe two different types of regions in L($\mathbf{r}$,V) maps: one characterized by unidirectional modulations, and the other where these modulations are notably absent (Fig. 1(e)). To illustrate the morphological difference between these areas, we show a domain boundary in Fig. 1(c,e). The transition is atomically smooth, and it does not show any obvious structural imperfections or surface buckling. The spatial extent of modulated regions can be as large as several micrometer size in different $x$~0.45 samples (Supplementary Information 3). We note that the modulations are not immediately obvious from STM topographs, which look indistinguishable in regions with or without the modulations, with no apparent change in Se:Te ratio or excess Fe concentration (Fig. 1(c), Supplementary Information 1).

To characterize the modulated regions, we acquire L($\mathbf{r}$,V) maps as a function of energy over a larger field-of-view (Fig. 2). Our first observation is that the modulation wavelength strongly changes with energy over a few meV energy range across the Fermi level. This trend is clear in Fourier transforms of L($\mathbf{r}$,V) maps, which reveal an arc-shaped wave vector oriented along one of the Fe-Fe lattice directions (**a**-axis) in momentum space. Notably, we observe almost no intensity along the other Fe-Fe direction (**b**-axis). The magnitude of this wave vector quickly decreases with increased energy (Fig. 2(f-h)).

Dispersing electronic modulations in differential conductance maps typically originate due to interference and scattering of electrons at the surface, also known as quasiparticle interference (QPI) imaging. For guidance in determining the origin of the dominant wave vector in our data, we refer to the band structure schematic of Fe(Te,Se) near $x$~0.45 (relevant bands shown in Fig. 2(a,b)). The Fermi surface of Fe(Te,Se) consists of hole pockets at Γ, and electron pockets at the M point [17,29]. By comparing the magnitude of the scattering vector and its dispersion velocity to the expected band structure from angle-resolved photoemission measurements [17,28,30], we can trace it to the scattering within the Γ hole pocket composed of $d_{xz}$ and $d_{yz}$ orbitals (Fig. 2(a,b)). Intra-pocket scattering at the M point can be ruled out based on: (1) the sign of the dispersion velocity that is opposite of what we observe, and (2) the empirical evidence that the electronic states near the Brillouin zone edge in these systems are difficult to be picked up by STM tips [20,22]. The strongly anisotropic scattering signature is consistent with scattering between $d_{yz}$ orbitals ($\mathbf{Q_a}$), while the scattering between the $d_{xz}$ orbitals ($\mathbf{Q_b}$) is much weaker and cannot be resolved in our measurements. The striking difference in the scattering strengths can be understood in terms of the variation of the spectral weight of different orbitals, giving rise to orbital-selective quasiparticles in the electronic nematic state [20]. This is also likely accompanied by the elongation of the Fermi surface, creating near-parallel Fermi sheets that enhance scattering along the **a**-axis (Fig. 2(a)). In FeSe, it was found that the spectral weight of quasiparticles associated with $d_{yz}$ orbitals is much larger than that of either $d_{xz}$ and $d_{xy}$ orbitals [20]. This picture is consistent with our spectroscopic measurements of modulated electronic regions of Fe(Te,Se), and point towards the emergence of local electronic nematic order in Fe(Te,Se) at $x$~0.45.

We observed local electronic nematicity in ~10 different $x$~0.45 Fe(Te,Se) single crystals, grown by three different research groups (Table S2), which highlights the ubiquity of the phenomenon, and allows us to explore the relationship between local nematicity and superconductivity. We find that electronic nematicity is only present in regions where superconducting coherence is suppressed (Fig. 3(f-j)), or even completely absent (Fig. 3(a-e)) at our measurement temperature of ~4.5 K. This is in contrast to areas where $C_2$-symmetric electronic modulations associated with electronic nematicity are notably absent, and superconducting gap is evident at the same temperature (Fig. 3(k-o)). While our observations demonstrate that superconductivity and static electronic nematicity in Fe(Te,Se) can co-exist, they also suggest a direct local competition between the two in this system. We note that QPI dispersions across different electronic nematic regions are comparable, regardless of the presence or absence of superconductivity. The scattering wave vectors at the Fermi level (~0.2 A$^{-1}$ in Fig. 3(d,i)) match well with ARPES measurements of the associated Γ pocket [30]. We also note that we are able to find nematic domains oriented along both inequivalent Fe-Fe lattice directions in the same sample (Supplementary Information 3), ruling out tip anisotropy as the cause of our observations.

Next, we explore why electronic nematic order may emerge in some regions, but not in others. We rule out the distribution of excess Fe interstitials, known to strongly affect electronic properties of Fe(Te,Se) [31–33], because their concentration in our samples is minimal and their individual positions are not correlated with electronic nematic regions (Supplementary Information 1). We also rule out variations in Se:Te composition by calculating local composition from STM topographs (Fig. S2, Supplementary Information 1). This leaves the possibility of structural inhomogeneity. While our samples are attached to sample plates using a standard procedure, and are not intentionally strained (Methods), small strain of a few tenths of a percent may be imparted on the sample in this geometry due to small differences in thermal contraction coefficients [18]. To investigate this, we employ the strain analysis established in our previous work, which can detect local variations in the atomic lattice constant with a fraction of a percent resolution [34,35] (Supplementary Information 2).

To examine the effects of external strain on Fe(Te,Se), we focus on a region with pronounced topographic variations (Fig. 4). By inspecting different components of the strain tensor, we reveal the existence of spatially varying strain, with the most pronounced magnitude along the **a**-axis (Fig. S3). By subtracting the strain components along **a**- and **b**-axes, we create the map of relative antisymmetric (anisotropic) strain, which shows spatial variations at the order of ~1% (Fig. 4(c)). We then compare the calculated antisymmetric strain with the local amplitude of electronic modulations over the same region of the sample (Fig. 4(d)). The two observables exhibit a remarkably high cross-correlation coefficient of ~0.6 (Fig. 4(g)), thus demonstrating that antisymmetric tensile strain can drive the emergence of local electronic nematic order. Interestingly, superconductivity is also strongly suppressed over the same strained regions (Fig. 4(f,h)), consistent with our observations in Fig. 3. This can be best visualized by comparing the average d$I$/d$V$ spectra in regions of different strain (Fig. 4(f)), where antisymmetric strain of ~1% magnitude accompanies a strong suppression of the coherence peaks and spectral filling of the gap. Cross-correlation between the relative coherence peak height (RCPH(**r**)) map and antisymmetric strain map also exhibits high anti-correlation with the coefficient -0.5 (Fig. 4(h)). In contrast however, the gap size $\Delta(\mathbf{r})$ does not evolve with strain, as coherence peaks remain at approximately the same energy (Fig. S8). Previous experiment hinted at the competition between nematic excitations and spectral gap size $\Delta(\mathbf{r})$ [10]. However, our observations demonstrate that the suppression of superconductivity in electronic nematic regions in Fe(Te,Se) is dominated by the loss of coherence, not the decrease in pairing strength as the static nematicity strengthens. This could in turn also suggest a reduced superfluid density in the same areas [6]. While small variations in the coherence peak height (d$I$/d$V$(**r**, ±$\Delta$)) have been detected directly on top of Se vs Te atoms [36,37], a nearly complete suppression of gap edge peaks and filling of the gap observed in average d$I$/d$V$ spectra in our work clearly go beyond this effect. This is further confirmed by calculating local chemical composition $x$, which shows no significant correlation with spatial variations in RCPH(**r**) (Fig. S10). Thus, inhomogeneity in local composition $x$ cannot be the dominant driver behind the spectral gap suppression.

Lastly, we compare the effects of strain near the nematic transition at $x$~0.45 and away from it. In contrast to our observations at $x$~0.45, we find that anisotropic strain in $x$~0.35 samples does not give rise to unidirectional scattering in differential conductance maps within our resolution (Fig. S9). Moreover, it leads to a significantly smaller suppression of RCPH for a comparable amount of anisotropic strain (Fig. S8). This is consistent with physical picture portrayed by bulk elastotransport measurements of a related superconductor Ba(Fe$_{1-x}$Co$_x$)$_2$As$_2$, where the suppression of superconducting $T_c$ by anisotropic strain was found to be strongly dependent on bulk composition and the most pronounced near the nematic critical point [38]. Our experiments provide a complementary insight by further demonstrating that this suppression is accompanied by strengthening of static electronic nematicity.

Our experiments shed light on the nematic transition in Fe(Te,Se), and reveal the emergence of electronic nematicity in microscopic regions around $x$~0.45, at a few Kelvin above zero temperature. These measurements alone cannot rule out the possibility that the transition is discontinuous without critical singularities, and that the emergence of nematic domains is associated with the rounding of a first-order transition by quenched disorder [39]. However, given the reported quantum critical fluctuations associated with a divergent nematic susceptibility in the same material [14], our findings are consistent with the existence of an underlying nematic quantum critical point in Fe(Te,Se) at absolute zero. In this scenario, the local static nematicity observed at $x\sim0.45$ could be a reflection of nematic critical fluctuations that are pinned to become static in local regions.

Recently, suppressed superconducting coherence and significant zero-energy spectral weight have been reported along topographic ribbons in Fe(Te,Se), which was attributed to a one-dimensional Majorana mode [7]. By applying our strain analysis to the data from Ref. [7], we find anisotropic local strain in these regions as high as ~3-5% (Supplementary Information 6). This is significantly larger than that in our data where we already observe significant suppression of superconducting coherence peaks and an increase in in-gap conductance. This strain analysis, in combination with our experiments, provides strong evidence that strain can be responsible for the suppression of superconducting coherence and gap filling observed in these regions. It remains to be seen to what extent any additional spectral weight due to a Majorana mode can exist within the same regions.

**Methods:**

$FeTe_{1-x}Se_x$ bulk single crystals were grown using the self-flux method, and post-annealed to remove excess Fe impurities. The Se:Te ratio estimated from STM topographs (Supplementary Information 1) is comparable to the expected value from energy-dispersive spectroscopy measurements. The samples are attached to the STM sample plate (Titanium) using H20E silver epoxy (EPO-TEK), and heated to ~100-150 °C to cure the epoxy. Aluminum cleave bar is attached to the top of the sample using the same type of the epoxy and cured at the same ~100-150 °C temperature. The samples are cleaved in UHV ($10^{-10}$ Torr pressure) to expose a clean surface free of contaminants, and immediately inserted into the STM. The cleave temperature used was either ~300 K or ~80 K; there was no noticeable difference in the quality of the surface between the two. We note that ~10 Fe(Te,Se) samples with $x\sim0.45$ have been measured in this work, and we show representative data from eight samples in the main text and Supplementary Information. For each composition studied, samples are labeled sequentially starting with number 1 (i.e. $x\sim0.45$ (samples 1-8) and $x\sim0.35$ (samples 1-2)). Chemical composition reported for each sample is the bulk value determined by energy dispersive X-ray spectroscopy, which is comparable to the composition determined by counting individual atoms in STM topographs (Table S1, Supplementary Information 1).

STM data was acquired using a Unisoku USM1300 STM at the base temperature of ~ 4.5 K. Spectroscopic measurements were made using a standard lock-in technique at 915 Hz frequency, with bias excitation as detailed in figure captions. STM tips used were home-made chemically-etched tungsten tips, annealed in UHV to bright orange color prior to STM imaging. We use normalized conductance L(**r**,V) maps (defined as dI/dV(**r**,V)/(I(**r**,V)/V)) in our analysis to remove the effect of the STM tip setup condition [20,22].

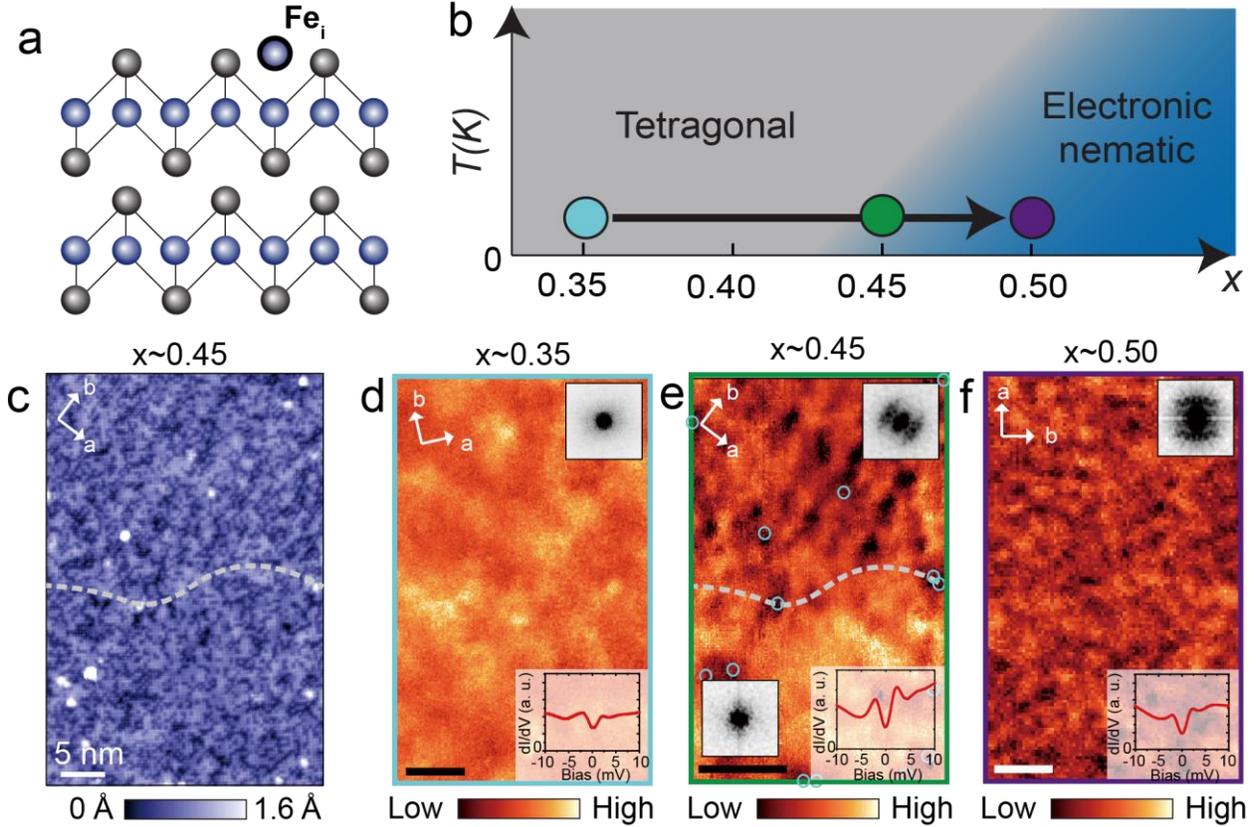

**Figure 1.** Nematic transition as a function of composition in Fe(Te,Se). (a) The schematic of crystal structure of Fe(Te,Se) with Fe (Se/Te) denoted as blue (gray) spheres. An example of excess interstitial Fe atom (Fe$_i$) located between 4 adjacent Se/Te atoms is circled in black line. (b) Temperature vs. Se composition (*x*) phase diagram portraying the electronic nematic transition. The three different compositions studied in this work are denoted by cyan (*x*~0.35), green (*x*~0.45) and purple circles (x~0.5). (c) STM topograph of a domain boundary between the four-fold symmetric region (bottom half) and a region hosting additional electronic modulations (top half), and (e) simultaneously acquired L(**r**,V) maps acquired at -2 mV on *x*~0.45 (sample 1). (d, f) L(**r**,V) maps acquired on top of *x*~0.35 and *x*~0.5 samples, respectively. Insets in lower right of (d-f) are associated spatially averaged *dI/dV*(**r**,V) spectra; insets in upper right of (d-f) are two fold-symmetrized Fourier transforms of corresponding L(**r**,V) maps. Excess interstitial Fe atoms seen as bright protrusions in STM topographs are circled in cyan in (e). Scale bars in (d-f) represent 10 nm distance. STM setup conditions: (c, e) $V_{sample}$ = -10 mV, $I_{set}$ = 100pA, $V_{exc}$ =2 mV. Inset spectrum: $V_{sample}$ = 10 mV, $I_{set}$ = 60pA, $V_{exc}$ =0.3 mV; (d) $V_{sample}$ = 10 mV, $I_{set}$ = 50pA, $V_{exc}$ =0.3 mV. (f) $V_{sample}$ = 10 mV, $I_{set}$ = 60pA, $V_{exc}$ = 0.3 mV.

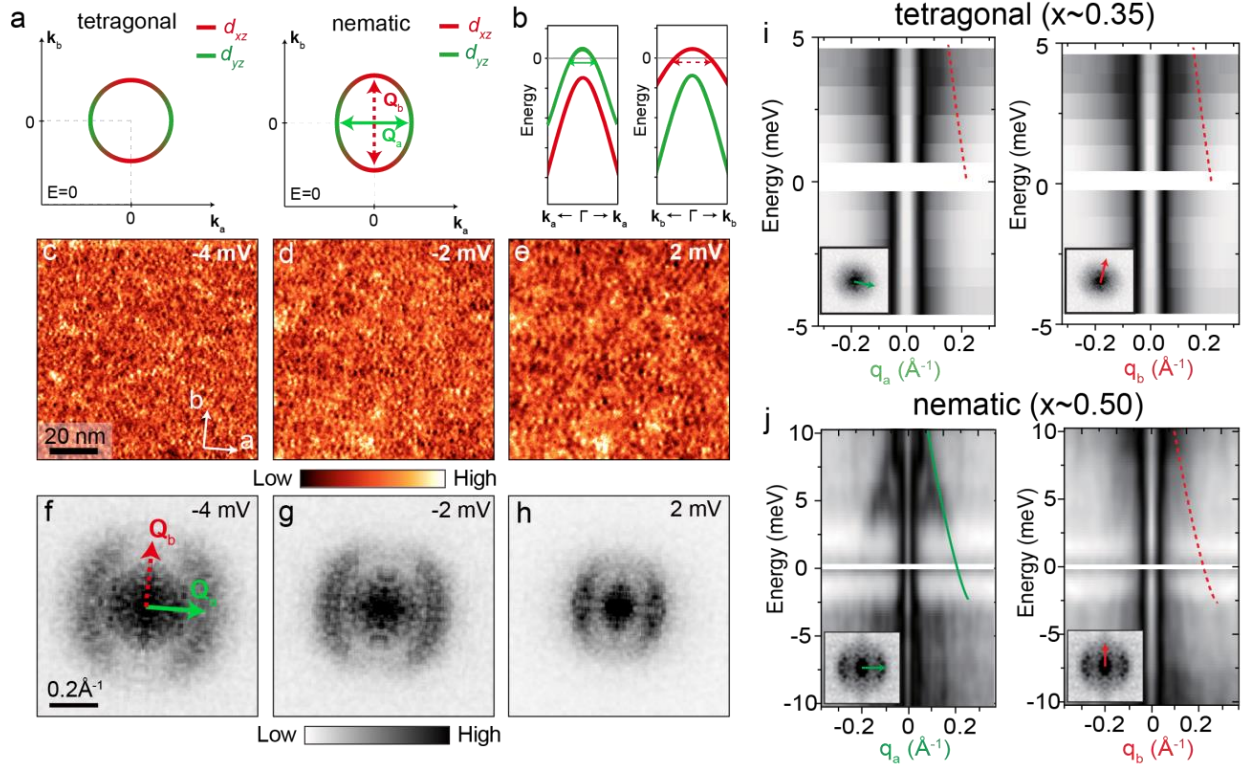

**Figure 2.** Visualizing anisotropy in electron scattering. (a) Schematic of relevant band in Fe(Te,Se) at Fermi level around Γ point. Orbital content of different electronic states is denoted by red ($d_{xz}$) and green ($d_{yz}$). The left panel represents the isotropic fermi surface of tetragonal Fe(Te,Se) sample while the right one is a schematic of an elongated Fermi surface in the nematic state, where quasiparticle interference (QPI) scattering appears along the longer parallel sheets ($\mathbf{Q_a}$), but not in the orthogonal direction ($\mathbf{Q_b}$). (b) The dispersion of two electronic bands around Γ along $\mathbf{k_a}$ and $\mathbf{k_b}$ directions. For simplicity, we omit drawing the $d_{xy}$ band in (b) that is pushed below the Fermi level. (c-e) Normalized differential conductance L(**r**,V) maps and (f-h) associated Fourier transforms (FTs) in Fe(Te,Se) with $x$~0.45, sample 2. The arrows in (f) mark the position of the expected scattering vectors on the inner Γ pocket shown in (a,b). For visual purposes, L(**r**,V) maps have been box-car-averaged with 1 pixel radius. FTs in (f-h) have also been two-fold symmetrized with respect to the **a**-axis. (i, j) Linecuts in two-fold symmetrized Fourier transforms of L(**r**,V) maps along $\mathbf{q_a}$ and $\mathbf{q_b}$ direction in $x$~0.35 (sample 2) and $x$~0.5 samples, respectively. Insets in (i,j) portray the directions along which the linecuts were taken in reciprocal space. Green line in (j) is a visual guide showing the dispersion of $\mathbf{Q_a}$, while the dashed red lines in (i,j) denote the absence of scattering vectors at equivalent positions. STM setup condition: (c-h) $V_{sample}$ = 5 mV, $I_{set}$ = 50pA, $V_{exc}$ =0.3 mV. (i) $V_{sample}$ = 5 mV, $I_{set}$ = 40pA, $V_{exc}$ =0.4 mV. (j) $V_{sample}$ = 10 mV, $I_{set}$ = 50pA, $V_{exc}$ =0.3 mV.

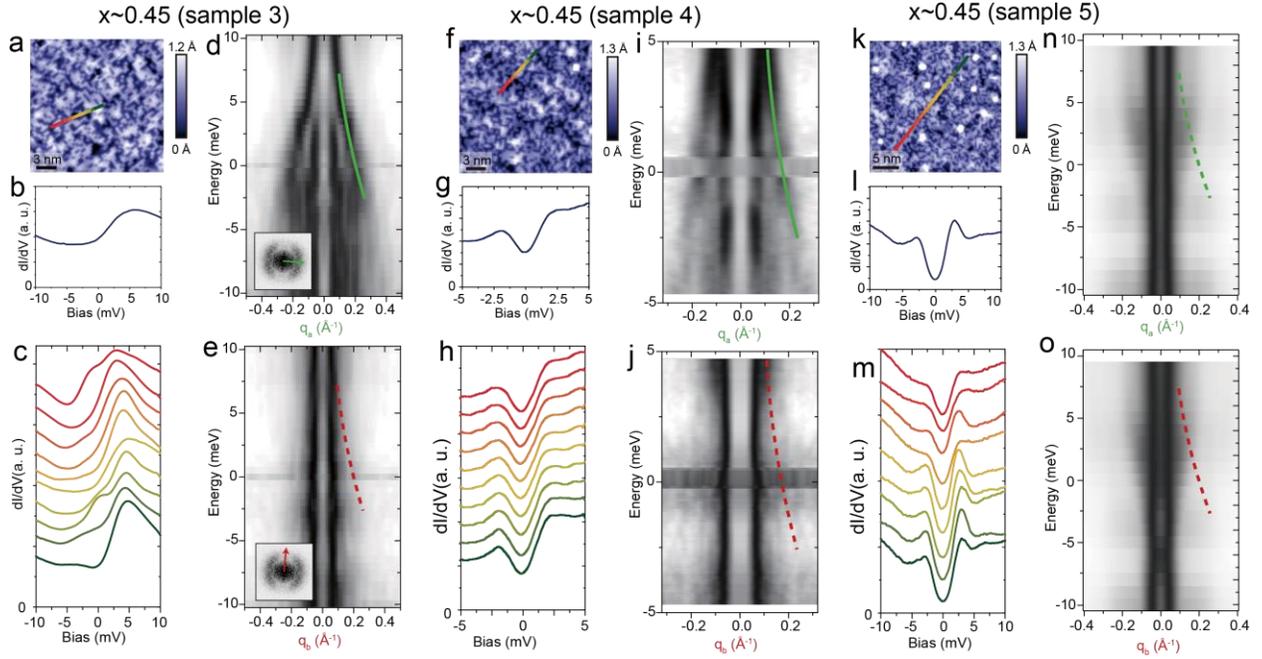

**Figure 3.** Spectroscopic-imaging scanning tunneling microscopy of critical composition. (a) STM topograph, (b) average d$I$/d$V$ spectrum, (c) linecuts of d$I$/d$V$ spectra and (d,e) linecuts in two-fold symmetrized Fourier transforms of L(**r**,V) maps along (d) $\mathbf{q}_a$ and (e) $\mathbf{q}_b$ direction in $x$~0.45, sample 3. Insets in (d,e) portray the directions along which the linecuts were taken in reciprocal space. (f-j) and (k-o) are equivalent panels corresponding to (a-e) for $x$~0.45 samples labeled 4 and 5, respectively. Green lines in (d,i) are visual guides showing the dispersion of $\mathbf{Q}_a$. Dashed lines in (e,j,n,o) denote the absence of scattering vectors expected at these positions. Each linecut was averaged by 7 consecutive pixels along transverse direction. STM setup condition: (a-e) Topograph: $V_{sample}$ = 400 mV, $I_{set}$ = 60pA; Spectrum and dispersion: $V_{sample}$ = 10 mV, $I_{set}$ = 60pA, $V_{exc}$ = 0.3 mV. (f-j) Topograph: $V_{sample}$ = 10 mV, $I_{set}$ = 60pA; Spectrum and dispersion: $V_{sample}$ = 5 mV, $I_{set}$ = 30pA, $V_{exc}$ = 0.2 mV; (k-o) Topograph: $V_{sample}$ = 2 mV, $I_{set}$ = 40pA; Spectrum and dispersion: $V_{sample}$ = -10 mV, $I_{set}$ = 100pA, $V_{exc}$ = 0.2 mV.

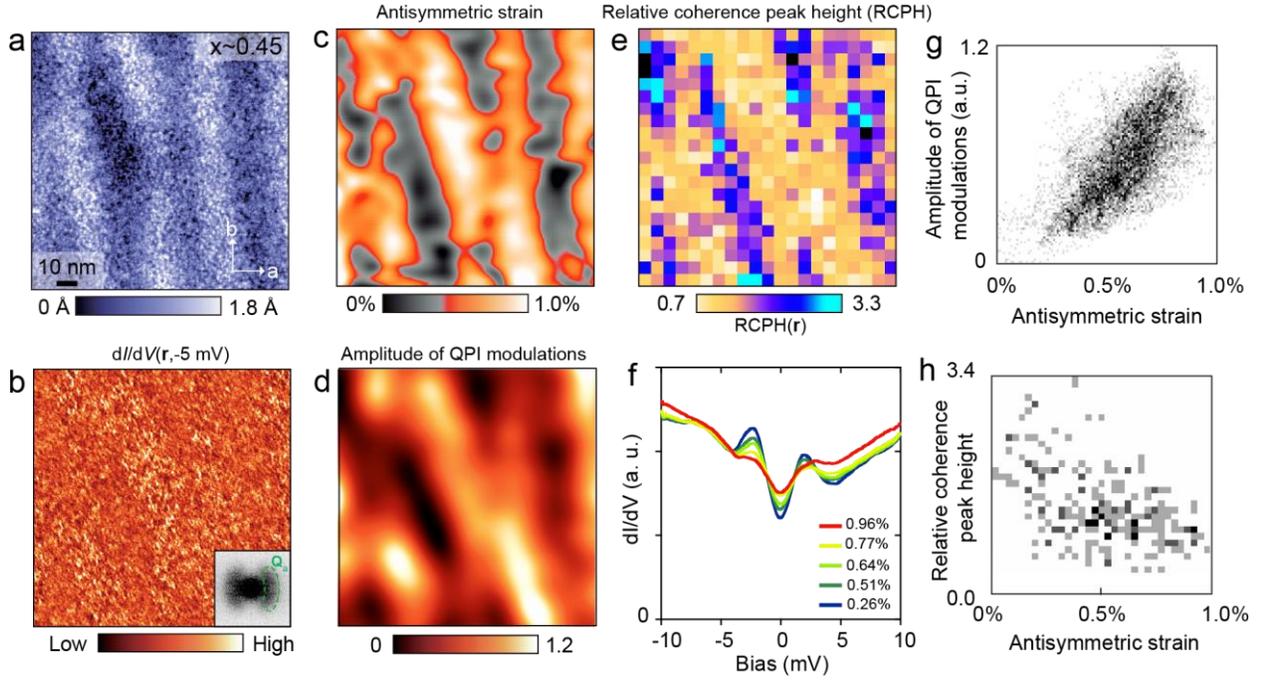

**Figure 4.** Interplay of strain, QPI amplitude and superconductivity at the nanoscale. (a) STM topograph, (b) d$I$/d$V$(**r**,-5 mV) map showing spatially inhomogeneous QPI signal, (c) antisymmetric strain map, (d) local amplitude of QPI from panel (b), and (e) average coherence peak height normalized by the zero-bias conductance (RCPH(**r**)=(d$I$/d$V$(**r**,$\Delta_+$) + d$I$/d$V$(**r**,$\Delta_-$))/(2·d$I$/d$V$(**r**,0))), all acquired over an identical area of Fe(Te,Se) ($x$~0.45, sample 6). Inset in (b) shows a two-fold symmetrized Fourier transform of the map in (b). Panel (d) was obtained by isolating the **Q**$_a$ peak in Fourier space (denoted by dashed green line in the inset of (b)) to create a Fourier-filtered image, and plotting the local amplitude of modulations extracted from the Fourier-filtered image. (f) Average d$I$/d$V$ spectra binned by the magnitude of strain in (c). (g) Correlation histogram between anisotropic strain in (c) and amplitude of QPI modulations in (d) showing a high cross-correlation coefficient of ~0.6. (h) Correlation histogram between RCPH and antisymmetric strain, showing high anti-correlation with coefficient -0.5 (formula used for cross-correlation can be found in Supplementary Information 4). STM setup condition: (a,b) $V_{sample}$ = -5 mV, $I_{set}$ = 10 pA, $V_{exc}$ =1 mV.

**Acknowledgements**

I.Z. gratefully acknowledges the support from Army Research Office Grant No. W911NF-17-1-0399 (STM experiments) and National Science Foundation Grant No. NSF-DMR-1654041 (strain analysis). The work at Brookhaven was supported by the Office of Basic Energy Sciences, U.S. Department of Energy (DOE) under Contract No. DE-SC0012704. Partial support was provided via the UC Santa Barbara NSF Quantum Foundry funded under the Q-AMASE-i initiative under award DMR-1906325 (S.D.W.). Z.W. acknowledges the support from U.S. Department of Energy, Basic Energy Sciences Grant No. DE-FG02-99ER45747. The work in Zhejiang University is supported by the National Key R&D Program of China under Grant No. 2016YFA0300402, and the National Natural Science Foundation of China (Grants No. NSFC-12074335 and No. 11974095).